\documentclass[11pt,aps,prc,superscriptaddress,preprintnumbers,showpacs,nofootinbib,endfloats]{revtex4}
\usepackage[dvips]{graphics,graphicx}
\usepackage{amsmath}
\begin{document}
\preprint{WU-HEP-02-3}
\preprint{TWC-02-1}
\title{Comparison of space-time evolutions of hot/dense matter in
$\sqrt{s_{NN}}$=17 and 130 GeV relativistic heavy ion collisions based
on a hydrodynamical model}
\author{Kenji Morita}
\email{morita@hep.phys.waseda.ac.jp}
\affiliation{Department of Physics, Waseda University, Tokyo 169-8555, Japan}
\author{Shin Muroya}
\email{muroya@yukawa.kyoto-u.ac.jp}
\affiliation{Tokuyama Women's College, Tokuyama, Yamaguchi 745-8511, Japan}
\author{Chiho Nonaka}
\email{nonaka@rarfaxp.riken.go.jp}
\affiliation{The Institute of Physical and Chemical Research (RIKEN), Wako, Saitama, 351-0198, Japan}
\author{Tetsufumi Hirano}
\email{hirano@nt.phys.s.u-tokyo.ac.jp}
\affiliation{Physics Department, University of Tokyo, Tokyo 113-0033, Japan}


\begin{abstract}
 Based on a hydrodynamical model, we compare 130 GeV/$A$ Au+Au
 collisions at RHIC and 17 GeV/$A$ Pb+Pb collisions at SPS. The model
 well reproduces the single-particle distributions of both RHIC and SPS.
 The numerical solution indicates that huge amount of collision energy
 in RHIC is mainly used to produce a large extent of hot fluid rather
 than to make a high temperature matter; longitudinal extent of the hot
 fluid in RHIC is much larger than that of SPS and initial energy
 density of the fluid is only 5\% higher than the one in SPS. The
 solution well describes the HBT radii at SPS energy but shows some
 deviations from the ones at RHIC.
\end{abstract}
\maketitle

\section{Introduction}\label{sec:intro}
One of the main purposes of ultra-relativistic heavy ion collision
experiments is to explore the property of the hot and dense matter
\cite{QM2001}. Recently the new experiment has begun to work at
BNL-RHIC of which higher collision energy than other experiments 
up to now provides us chances to produce a new state of matter,
quark-gluon plasma (QGP) with distinct possibility. However, the
complicated collision processes composed of multiparticle productions
and many-body interactions make it difficult to understand the
properties of the hot matter. Therefore, a simple
dynamical description of the system as a basis for the deeper
understanding is indispensable. 

 Relativistic hydrodynamical models are well-established
phenomenological tools for describing high energy nucleus-nucleus
collisions and subsequent multiparticle production
\cite{Bjorken,Gersdorff_PRD34,Akase_PTP85,Alam_PREP273,Ornik_PRC54,
Hung_PRC57,Kolb_PRC62}.
In this paper, we use a (3+1)-dimensional hydrodynamical model
\cite{Ishii_PRD46} with a QCD phase transition. We assume a cylindrical
symmetry to the collision dynamics. Thus, our discussion is limited to
the central collisions only. By virtue of the simple picture of our
model, we can easily analyze both SPS and RHIC data with the same
numerical code. Most of hydrodynamical calculations for RHIC data use
Bjorken's scaling solution \cite{Bjorken} for the longitudinal
direction. For example, Kolb \textit{et al.} analyzed hadronic tranverse
mass spectra and anisotropic flow \cite{Kolb_PLB500}. Zschiesche
\textit{et al.} \cite{Zschiesche_nucl7037} investigated the HBT
radii. These calculations assume the longitudinal boost-invariant infinite
source. Though recently some hybrid models are used
\cite{Bass_PRC61,Teaney_PRL86,Teaney_nucl0110037} for the description of
the hadronic phase, we here use a conventional description in which the
hadronic phase is in local equilibrium.

 In this paper, concentrating our discussion on the central collisions, 
we reproduce the single-partcle spectra of hadrons at the beginning. In
the hydrodynamical model, single-particle distributions are used as
inputs in order to determine initial parameters rather
than outputs. However, it is not trivial whether we succeed to reproduce
experiments with ``natural parameters'' or not.
Based on the solutions of hydrodynamical equations, we discuss the
physical properties and the space-time evolution of the fluids in SPS 
and RHIC. We also evaluate the two-pion correlation functions and
analyze the HBT radii. As a subsequent work of Ref.~\cite{Letter}, we focus
our discussion on comparison of the RHIC results and the SPS results based
on the same numerical code.

As is well-known, the two-particle correlation function gives us
information on the size of the particle source
\cite{Wiedemann_PREP,Weiner_PREP}.
In the cases of the relativisitic heavy ion collisions, the correlation
function tells us about the freeze-out which should be far
from the static source. Thus, dynamical models such as hydrodynamical
models are indispensable for understanding the relation between observed
correlation functions and the space-time history of the system. However, 
up to now, any dynamical model assuming QGP failed to explain the
experimental HBT radii in RHIC consistently with the single-particle
spectra \cite{Zschiesche_nucl7037,Soff_PRL} and elliptic flow
\cite{Heinz_hep0111075}, 
as being known as ``HBT puzzle''.\footnote{In
Ref.~\cite{Humanic_nucl0203004}, a hadronic rescattering model is shown
to reproduce these quantities.} 
We study the HBT radii in the framework of a hydrodynamical model which
takes account of both transverse and longitudinal flow appropriately
with a simple initial condition.

In the next section, we explain our model. In Sec.~\ref{sec:s-t_evo}, we
discuss the space-time evolution of the fluid. In Sec.~\ref{sec:hbt}, we
present the result of two-particle correlation. Section
~\ref{sec:conclusion} is devoted for the concluding remarks.

\section{Hydrodynamical Model}\label{sec:model}
Let the system achieve the local thermal and chemical equilibrium
shortly after a collision of two incident nuclei. This 
relaxation process cannot be described by the hydrodynamical model. 
The hydrodynamical model starts at initial time, $\tau_0$, at which
thermal and chemical equilibrium are established at least locally.
The hydrodynamical equations are given as
\begin{equation}
 \partial_\mu T^{\mu\nu}=0. \label{eq:hydro}
\end{equation}
We assume the perfect fluid for simplicity. Hence, energy-momentum
tensor is given as
\begin{equation}
 T^{\mu\nu}=(\epsilon +P)U^{\mu}U^{\nu}-Pg^{\mu\nu},
\end{equation}
with $U^\mu, \epsilon$ and $P$ being four velocities of a fluid element,
energy density and pressure, respectively. These are treated as local
quantities. We numerically solve the above equations
together with the net baryon number conservation law,
\begin{equation}
 \partial_\mu (n_{\text{B}}U^\mu) =0, \label{eq:baryon}
\end{equation}
where $n_{\text{B}}$ is the net baryon number density and is also
treated as a local quantity. Putting the $z$ axis as a collision axis,
we use a cylindrical coordinate system as follows;
\begin{align}
 t &= \tau \cosh\eta , \\
 x &= r \cos\phi , \\
 y &= r \sin\phi , \\
 z &= \tau \sinh\eta .
\end{align}
Focusing our discussion on central collisions, we may assume the
cylindrical symmetry on the system. Therefore, by virtue of an identity
$U_\mu U^\mu=1$, the four velocity can be expressed
by two rapidity-like varibles $Y_{\text{L}}$ and $Y_{\text{T}}$;
\begin{align}
 U^\tau &= \cosh(Y_{\text{L}}-\eta)\cosh Y_{\text{T}}, \\
 U^\eta &= \sinh(Y_{\text{L}}-\eta)\cosh Y_{\text{T}}, \\
 U^r &= \sinh Y_{\text{T}}.
\end{align}
Most of hydrodynamical calculations which analyze RHIC data use
Bjorken's scaling solution $Y_{\text{L}}=\eta$. Putting the solution as
an ansatz reduces numerical tasks very much but the analyses are limited
to midrapidity region only. We solve not only transverse expansion but
also the longitudinal expansion explicitly. Numerical procedure for
solving the coupled equations \eqref{eq:hydro} and \eqref{eq:baryon} is
explained in \cite{Ishii_PRD46}. In this algorhythm, we solve the
entropy and baryon number conservation law explicitly. Throughout our
calculation, the total energy, entropy and baryon number are conserved
within 5\% of accuracy at the time step $\delta \tau = 0.01$
fm/c. 

\begin{figure}[ht]
 \includegraphics[width=0.8\textwidth]{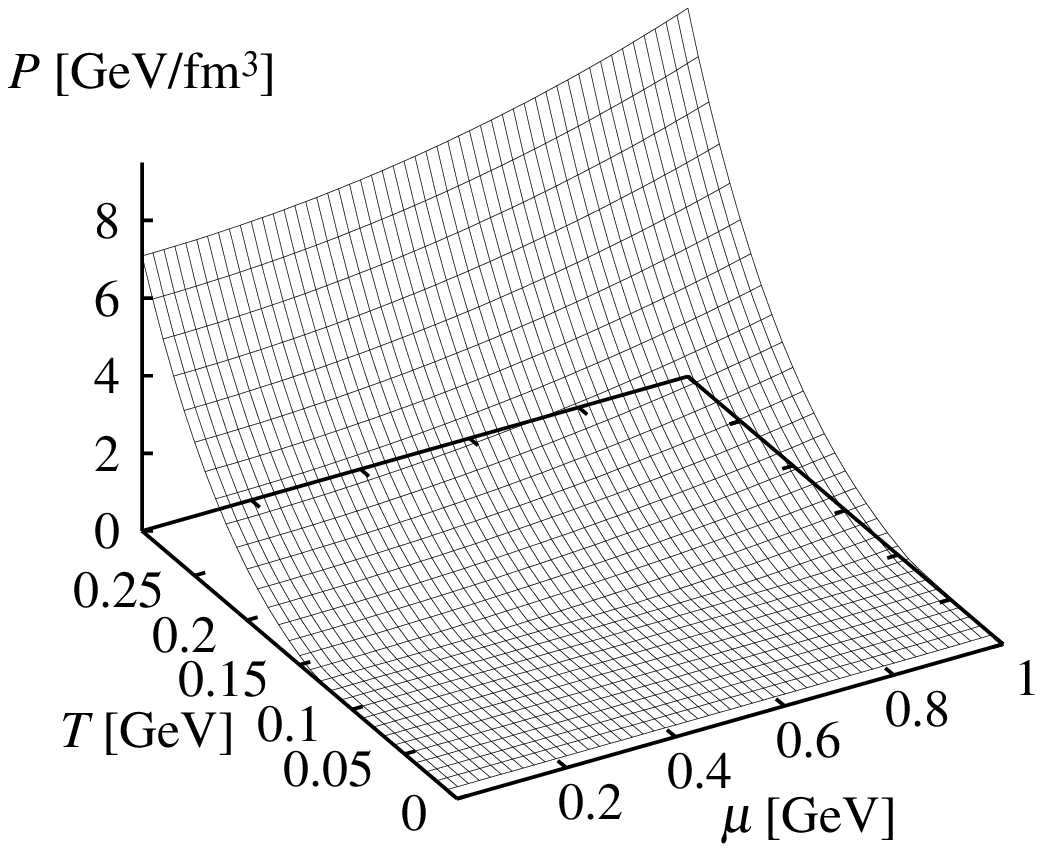}
 \caption{\label{fig:eos}Pressure $P(T,\mu_{\text{B}})$ distribution}
\end{figure}

In order to solve the hydrodynamical equations, we must fix the equation
of state (EoS). We adopt a bag model EoS with a first order phase
transition. The QGP phase is composed of a free gas of
massless u, d, s quarks and gluons. Hadronic phase is also assumed to be
a free gas but with excluded volume correction. All hadrons are included
up to 2 GeV/$c^2$ of mass except for hyperons. Putting the critical
temperature as $T_{\text{c}}=160$ MeV at vanishing baryon density, we
get the bag constant $B^{1/4}=233$ MeV. We display the pressure as a
function of temperature and baryonic chemical potential in
Fig.~\ref{fig:eos}. See \cite{Nonaka_EPJC} for the further detail of the
EoS and numerical treatment of the first order phase transition in
solving the hydrodynamical equations.

\begin{figure}[ht]
 \includegraphics[width=0.8\textwidth]{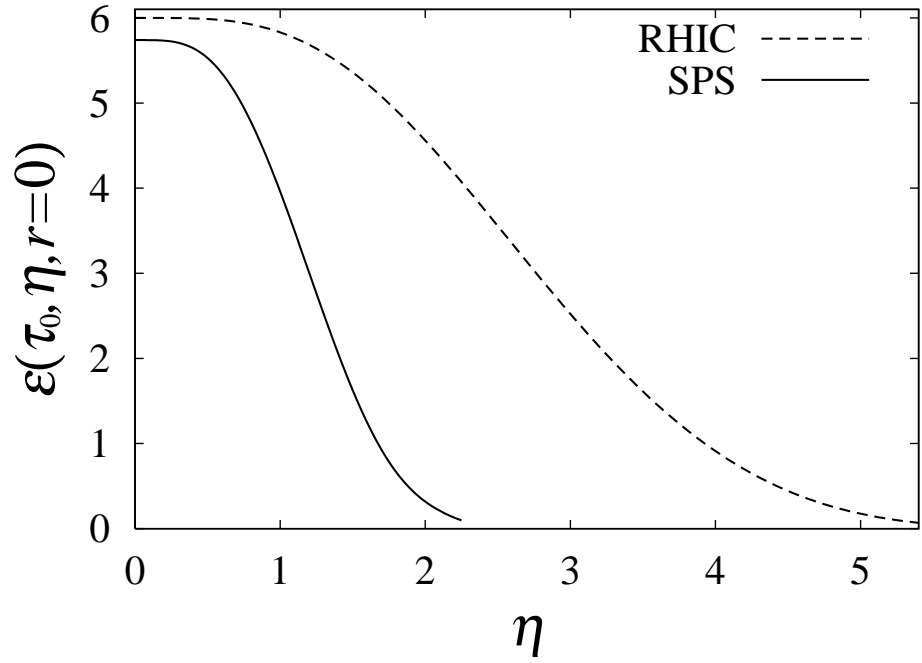}
 \caption{\label{fig:elong}Initial energy density distribution at $r=0$ in
 $\eta$-direction. Dashed line stands for the RHIC case while solid line
 stands for the SPS case. }
\end{figure}

\begin{figure}[ht]
 \includegraphics[width=0.8\textwidth]{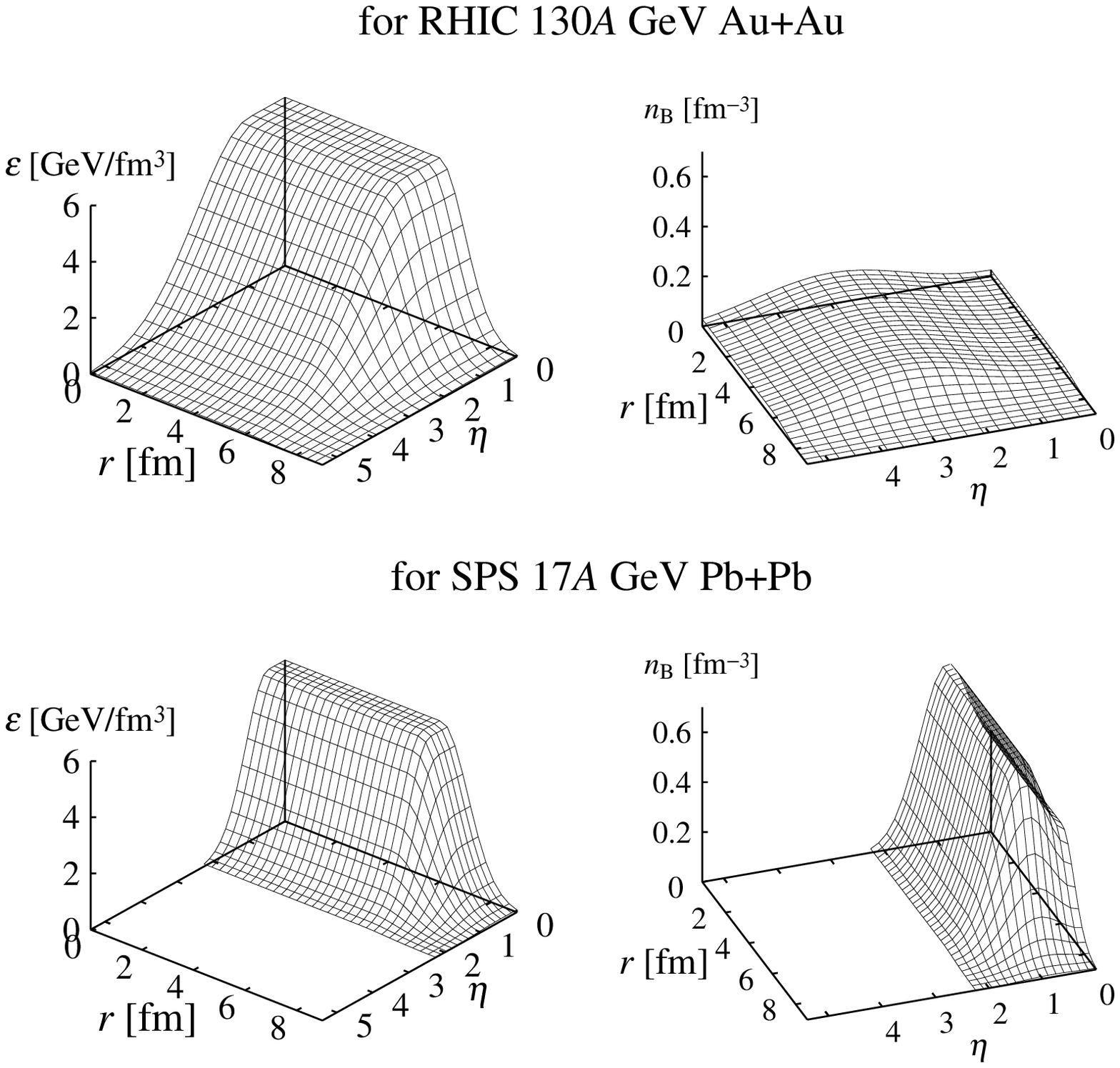}
 \caption{\label{fig:initial}Initial energy density (left) and net baryon
 number density (right) distribution for RHIC and SPS. }
\end{figure}

\begin{figure}[ht]
 \includegraphics[width=0.8\textwidth]{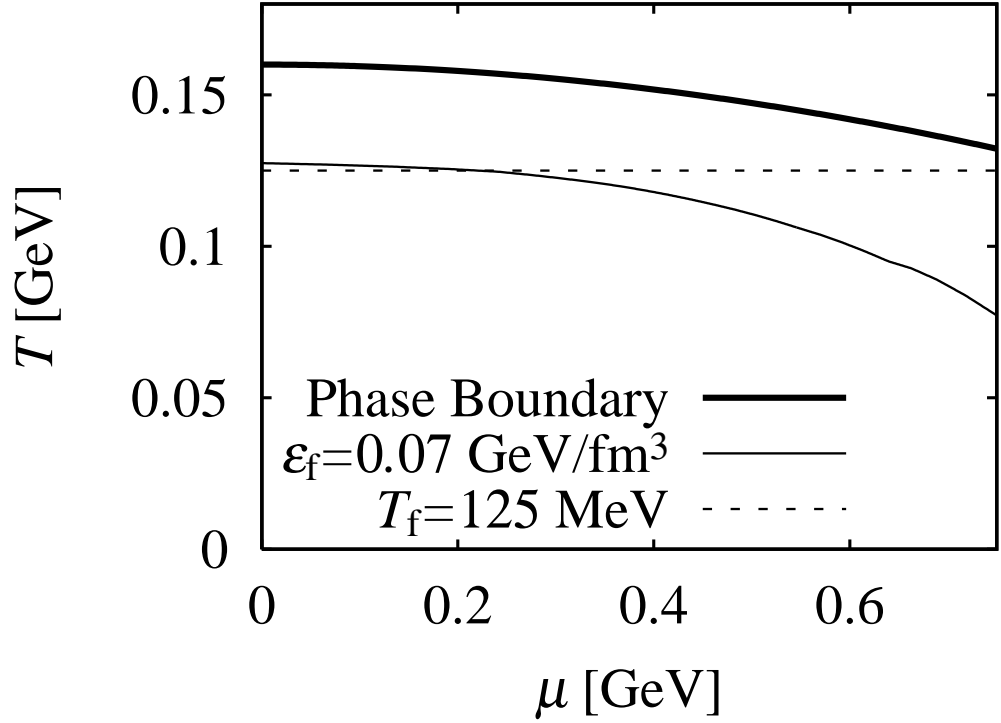}
 \caption{\label{fig:phase}Phase boundary (thick line) and freeze-out
 lines on $T-\mu_{\text{B}}$ plane. Thin solid line denotes the constant
 energy density contour as a freeze-out condition in SPS. Dashed line
 stands for constant temperature line, which is a freeze-out condition in
 RHIC.}
\end{figure}

We assume that the system achieves local equilibrium and begins to
expand hydrodynamically at $\tau=\tau_0=1.0$ fm/$c$. We put the initial
conditions on this hyperbola. Bjorken's scaling solution is used as the 
initial condition of the longitudinal flow.
Transverse flow is simply neglected at the initial time. We parameterize
the initial energy density distribution $\epsilon(\tau_0, \eta, r)$ and
net baryon number density distribution $n_{\text{B}}(\tau_0, \eta, r)$ as,
\begin{align}
 \epsilon(\tau_0,\eta,r) &= \epsilon_{\text{max}}
 \exp 
 \left[-\frac{(|\eta|-\eta_0)^2}{2\cdot \sigma_\eta^2}\theta (|\eta|-\eta_0)
 -\frac{(r-r_0)^2}{2\cdot \sigma_r^2}\theta (r-r_0)
 \right] ,  \label{eq:eini} \\
 n_{\text{B}}(\tau_0, \eta, r) &= n_{\text{B0}}
 \left\{ \exp
 \left[-\frac{(\eta-\eta_{\text{D}})^2}{2\cdot\sigma_{\text{D}}^2}\right]
 + \exp
 \left[-\frac{(\eta+\eta_{\text{D}})^2}{2\cdot\sigma_{\text{D}}^2}\right]
 \right\} \exp\left[-\frac{(r-r_0)^2}{2\cdot \sigma_r^2}\theta (r-r_0)
 \right] .
 \label{eq:nbini}
\end{align}
The energy density distribution of the longitudinal direction
\eqref{eq:eini} has a central plateau characterized by $\eta_0$ and a
gaussian tail whose width is given by $\sigma_\eta$
(Fig.~\ref{fig:elong}), while
the net baryon number distribution is a superposition of the two
gaussians of which peaks exist at $\pm\eta_{\text{D}}$. 
For the transverse direction, both are parametrized
by a flat region with gaussian smearing near the edge
(Fig.~\ref{fig:initial}). For a nucleus with mass number $A$, the
relation among these quantities is given by $\sigma_{\text{r}}+r_0=1.2
A^{1/3}$. \footnote{We adopt the initial condition as a natural and the
simplest extension of the (1+1)-dimensional Bjorken's picture
\cite{Bjorken} and as a basis for the further improvement. }
Once these parameters are fixed, we can solve the
hydrodynamical equations and pursue the space-time evolution of the
fluid. These initial parameters are so chosen that the model reproduces
the single-particle spectra measured in the experiments. The
single-particle spectra can be calculated by making use of
of the Cooper-Frye formula \cite{Cooper_Frye}
\begin{equation}
 E_{\mathbf{k}}\frac{dN_i}{d^3\mathbf{k}}=
  \frac{g_i}{(2\pi)^3} \int_{\Sigma}k_\mu d\sigma^\mu 
  \frac{1}{\exp[(U_\nu k^\nu - \mu_{\text{B}})/T_{\text{f}}]\mp 1} 
  \label{eq:c-f},
\end{equation}
where $g_i$ is a degeneracy of the hadrons and $T_{\text{f}}$ is a
freeze-out temperature. The sign is plus for fermions and minus for
bosons. Integration is performed on 3-dimensional freeze-out
hypersurface $\Sigma$. By virtue of the Lagrangian hydrodynamics, 
contribution from the time-like hypersurface is expected to be small and
the space-like hypersurface dominates the particle emission at
freeze-out; we employ the non-covariant prescription  $k_\mu d\sigma^\mu
\simeq k_\tau d\sigma^\tau$ for the sake of simplicity in the numerical
treatment. 
In this approximation, total counted energy evaluated
from Eq.\eqref{eq:c-f} is slightly larger than 90\% of the total energy of
the initial fluid; we regard the approximation works well enough. 
At SPS, we assume that the freeze-out occurs at a 
energy density $\epsilon_{\text{f}}$ and at a
temperature $T_{\text{f}}$, at the RHIC energy.
We also assume that the thermal and the chemical
freeze-out are taken to happen simultaneouly. We show the freeze-out
lines and the phase boundary on
$T-\mu_{\text{B}}$ plane in Fig.~\ref{fig:phase}. 
Note that two freeze-out lines in the figure do not differ
at low baryonic chemical potential (Fig.~\ref{fig:phase}). 

We take account of the particles emitted from
resonance decay as well as the direct emission from the freeze-out
hypersurface. We include the decay processes $\rho\rightarrow 2\pi$,
$\omega\rightarrow 3\pi$, $\eta\rightarrow 3\pi$, $K^* \rightarrow \pi
K$, and $\Delta\rightarrow N\pi$
\cite{Sollfrank_ZPHYS52,Hirano_PRL}. These resonances are also assumed to
be thermally emitted from the freeze-out hypersurface.

\begin{figure}[ht]
 \vspace*{-1mm}
 \includegraphics[width=0.75\textwidth]{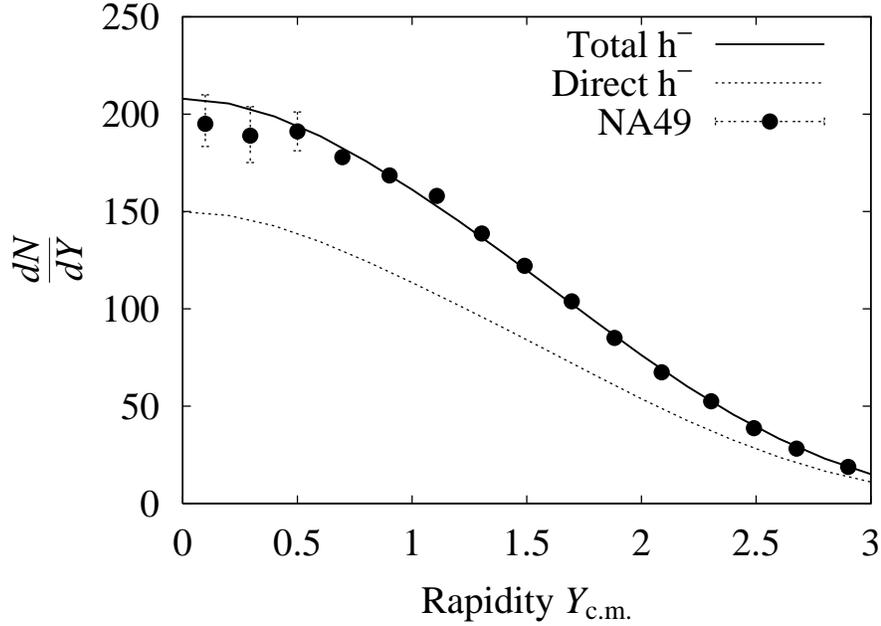}
 \caption{\label{fig:dndy_sps}Rapidity distribution of negatively
 charged hadrons in Pb+Pb collisions at SPS. Closed circles are
 experimental data which are taken from \cite{NA49_PRL82}. Solid and
 dashed lines stand for our result of total yield and contribution from
 direct particles.}
\end{figure}

\begin{figure}[ht]
 \includegraphics[width=0.8\textwidth]{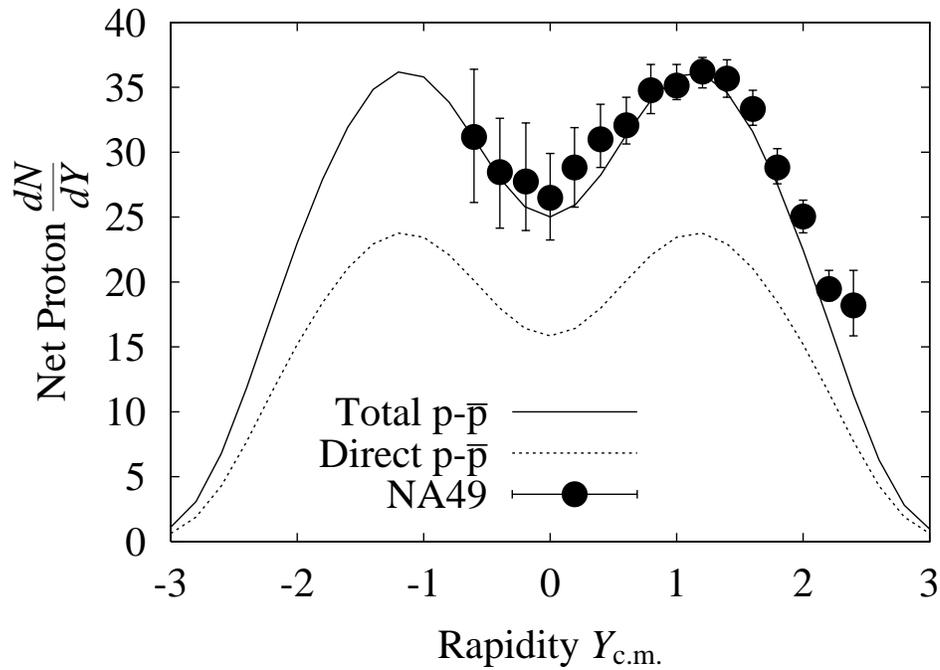}
 \caption{\label{fig:netproton_sps}Rapidity distribution of net protons
 in Pb+Pb collisions at SPS. Meanings of symbols and lines are the same
 as in Fig.~\ref{fig:dndy_sps}.}
\end{figure}

\begin{figure}[ht]
 \includegraphics[width=0.75\textwidth]{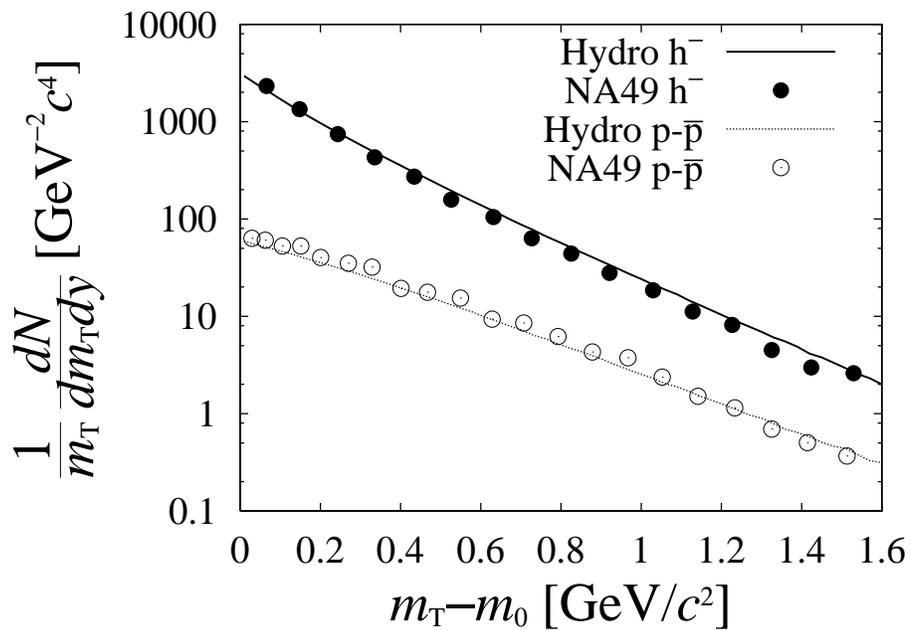}
 \caption{\label{fig:mt_sps}Transverse mass distributions of
 negatively charged hadrons and net protons in Pb+Pb collisions at SPS. 
 Closed and open circles are the experimental data of negatively charged
 hadrons and net protons, respectively. Solid and dashed lines stand for
 charged hadrons and net protons of our result, respectively.}
\end{figure}

\begin{figure}[ht]
 \includegraphics[width=0.75\textwidth]{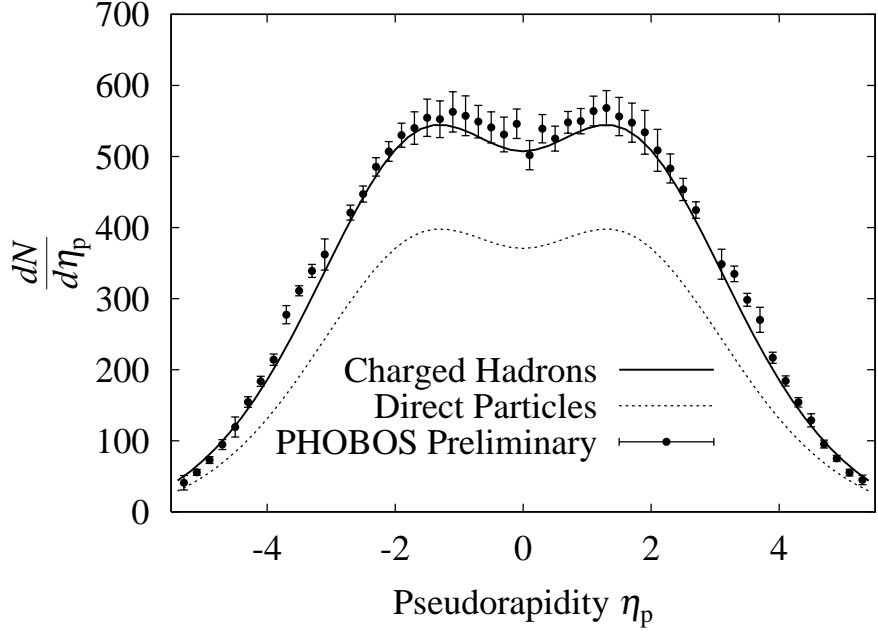}
 \caption{\label{fig:dndeta_rhic}Pseudorapidity $\eta_p$ distribution of
 charged hadrons. Solid line shows our result ($\pi, K, p$) including
 resonance contribution. Dotted line denotes contribution of the directly
 emitted particles from the freeze-out hypersurface. Closed circles are
 preliminary result from the PHOBOS Collaboration \cite{PHOBOS_dndeta}.}
\end{figure}

\begin{figure}[ht]
 \includegraphics[width=0.75\textwidth]{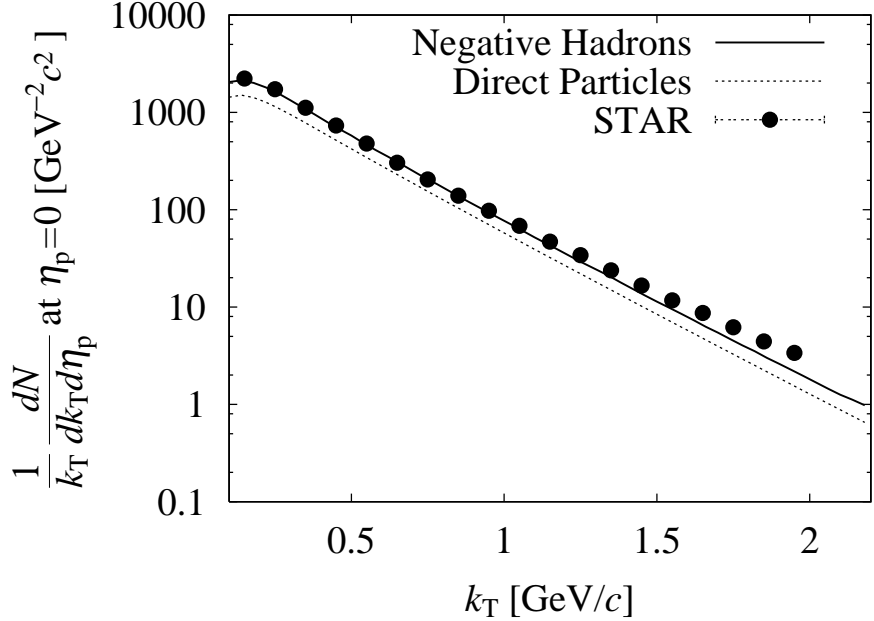}
 \caption{\label{fig:pt_rhic}Transverse momentum spectrum of negatively
 charged hadrons. As in Fig.\ \ref{fig:dndeta_rhic}, solid line and
 dotted line show total number of particles and directly emitted
 particles from the freeze-out hypersurface, respectively. Closed
 circles are data from the STAR Collaboration \cite{STAR_PRL87}.}
\end{figure}

\begin{figure}[ht]
 \includegraphics[width=0.75\textwidth]{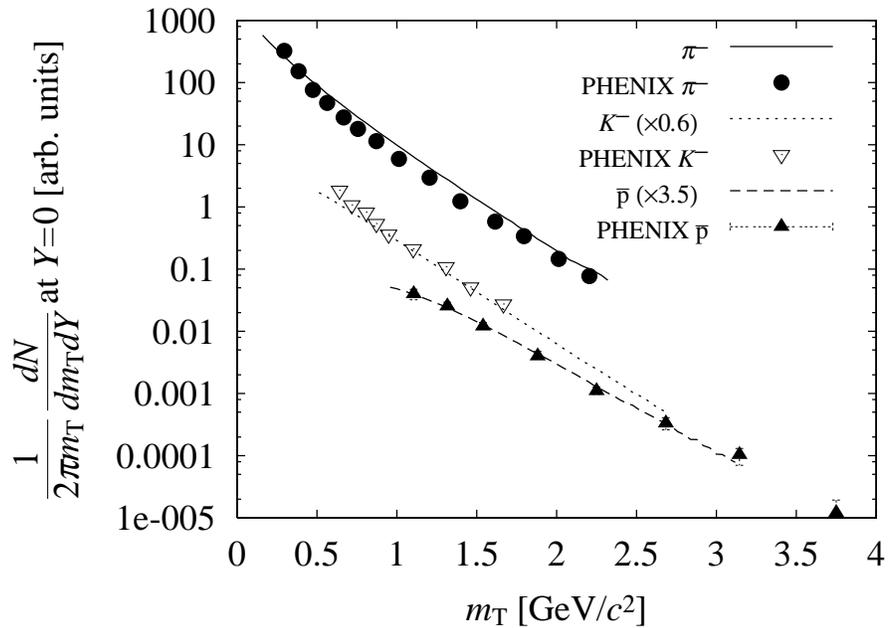}
 \caption{\label{fig:mt_rhic}Transverse mass spectra of negatively charged
 hadrons. Solid line, dotted line and dashed line denote $\pi^-$, $K^-$
 and $\bar{p}$ yield of our result. $K^-$ and $\bar{p}$ spectra are scaled
 down by factor 0.1 and 0.01, respectively. Closed circles, open triangles
 and closed triangles are preliminary data from the PHENIX Collaboration
 \cite{PHENIX_mt}. }
\end{figure}

\begin{figure}[ht]
 \includegraphics[width=0.75\textwidth]{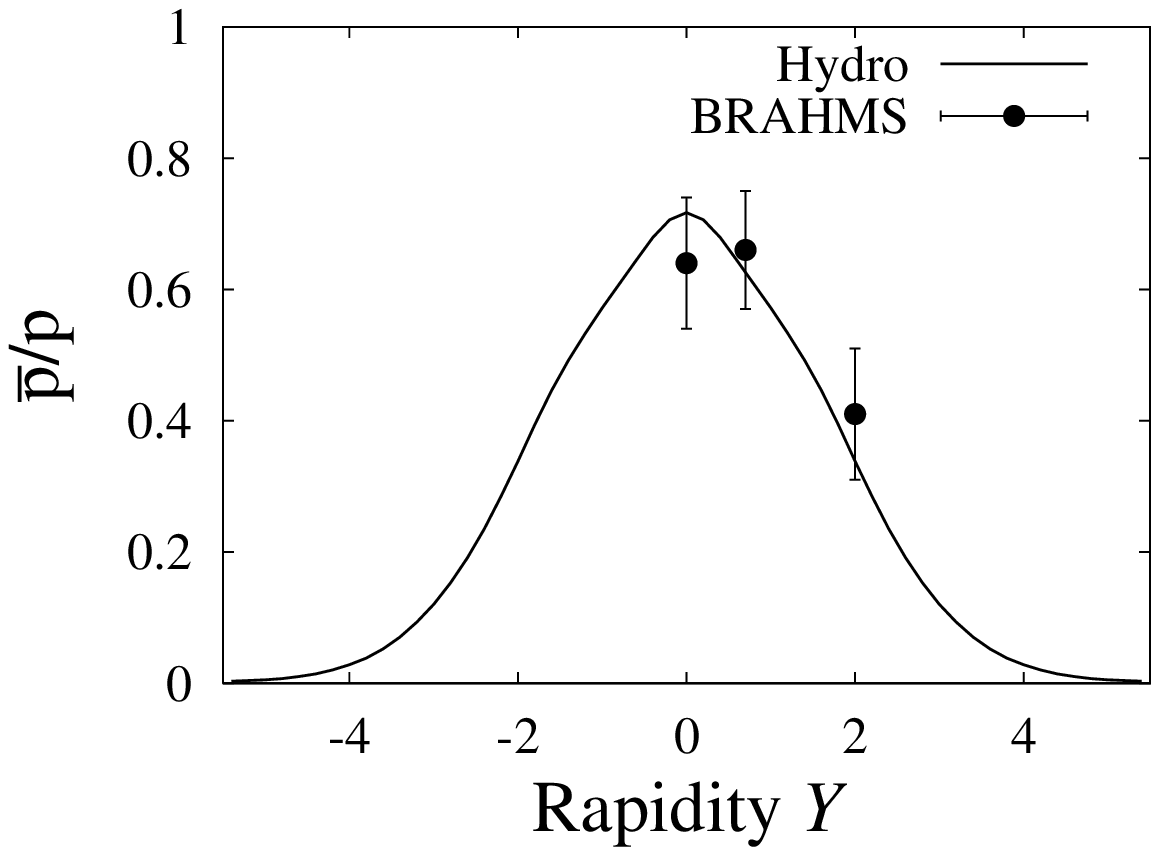}
 \caption{\label{fig:ratio_rhic}Rapidity dependence of anti-proton to proton
 ratio. Experimental data are taken from the BRAHMS Collaboration
 \cite{BRAHMS_PRL}.}
\end{figure}

Two sets of initial parameter are summarized in Table \ref{tbl:param}.
Figures \ref{fig:dndy_sps}-\ref{fig:mt_sps} show single-particle spectra
in 17 $A$GeV Pb+Pb collisions at SPS. Our model well reproduces the
experimental data with parameters in Table \ref{tbl:param}. Also in 130
$A$GeV Au+Au collisions, our model shows good agreement with the data as
in Figs.~\ref{fig:dndeta_rhic}-\ref{fig:ratio_rhic}. However, we note
that our model fails to produce enough number of anti-protons and
overestimates the kaon yield in Fig.~\ref{fig:mt_rhic}, where we
multiply factors of 0.6 for kaons and 3.5 for anti-protons for clear
comparison of the slopes \cite{Letter}. This discrepancy may indicate
the need for more sophisticated freeze-out mechanism.

\begin{table}
 \caption{\label{tbl:param}Initial parameter set.}
 \begin{ruledtabular}
  \begin{tabular}[t]{p{4cm}ll}
   & SPS Pb+Pb & RHIC Au+Au \\ \hline 
   Maximum initial energy density  $\epsilon_{\text{max}}$ & 5.74
   GeV/fm$^3$ & 6.0 GeV/fm$^3$ \\ 
   ``Maximum'' initial net baryon density  $n_{\text{B0}}$ & 0.7 fm$^{-3}$ &
   0.125 fm$^{-3}$ \\
   Longitudinal gaussian width $\sigma_{\eta}$ of initial energy density &
   0.61 & 1.47 \\
   Longitudinal extension $\eta_0$ of the flat region in the initial energy
   density & 0.48 & 1.0 \\
   Longitudinal gaussian width $\sigma_{\text{D}}$ of the initial net baryon
   density & 0.52 & 1.4 \\ 
   Space-time rapidity $\eta_{\text{D}}$ at maximum of the initial net baryon
   distribution & 0.82 & 3.0 \\
   Gaussian smearing parameter $\sigma_{\text{r}}$ of the transverse profile
   & 1.0 fm & 1.0 fm\\
   Freeze-out condition & $E_{\text{f}}=70$ MeV/fm$^3$ & $T_{\text{f}}$=125
   MeV \\
  \end{tabular}
 \end{ruledtabular}
\end{table}

\section{Space-time evolution}\label{sec:s-t_evo}

\begin{figure}[ht]
 \includegraphics[width=0.9\textwidth]{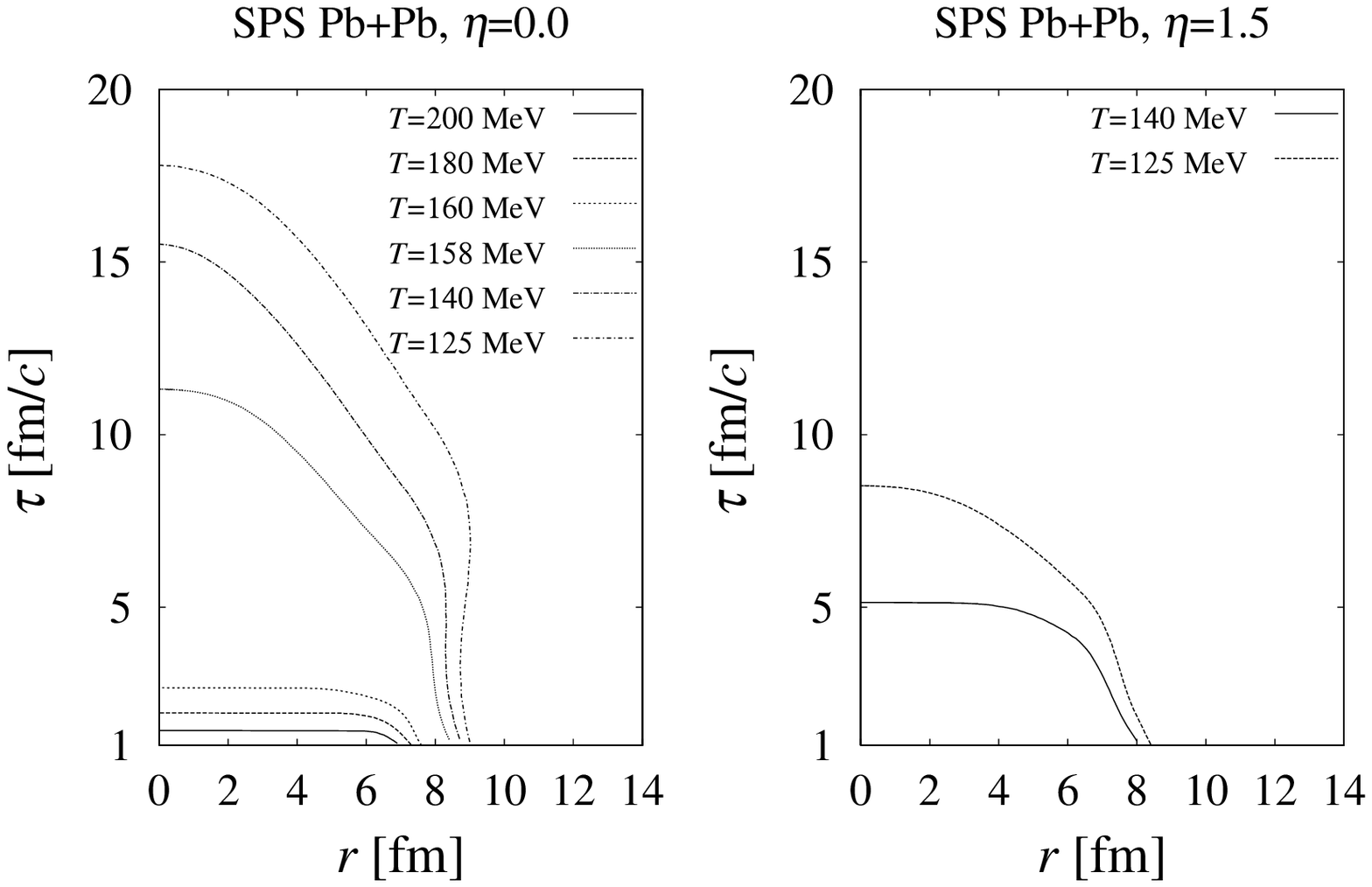}
 \caption{\label{fig:r-t_sps}Temperature contour on $r-\tau$ plane at
 SPS. Left: $\eta=0$ section. Right: $\eta=1.5$ section.}
\end{figure}

\begin{figure}[ht]
 \includegraphics[width=0.9\textwidth]{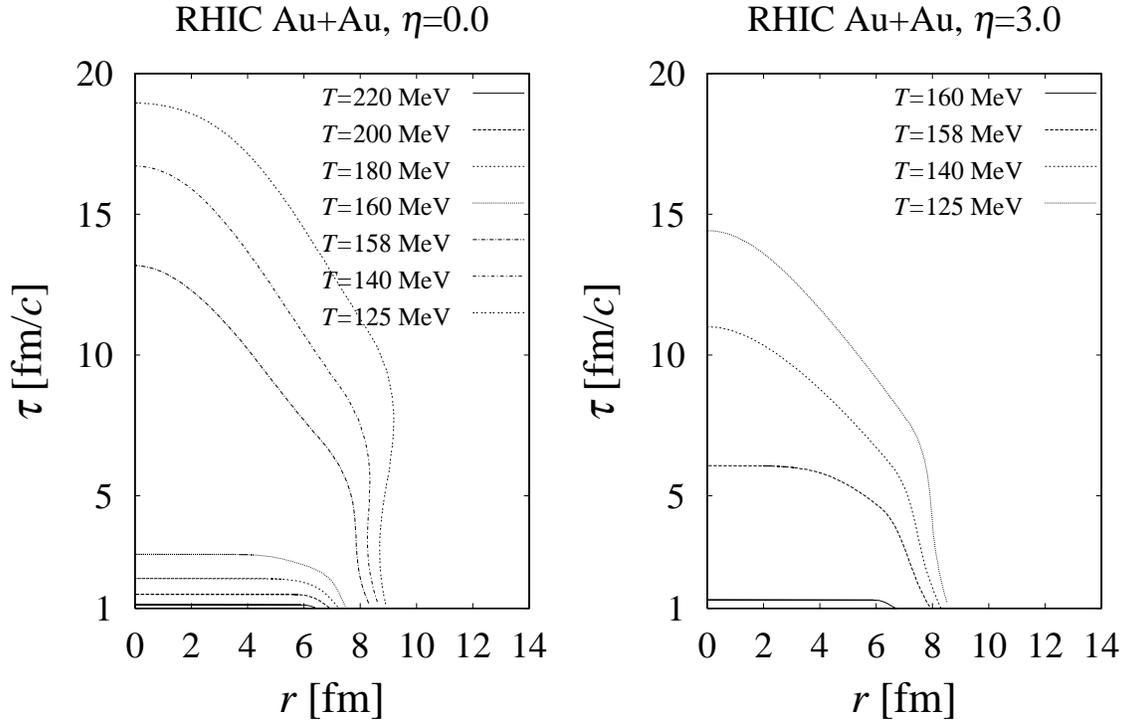}
 \caption{\label{fig:r-t_rhic}Temperature contour on $r-\tau$ plane at
 RHIC. Left: $\eta=0$ section. Right: $\eta=3.0$ section.}
\end{figure}

\begin{figure}[ht]
 \includegraphics[width=0.9\textwidth]{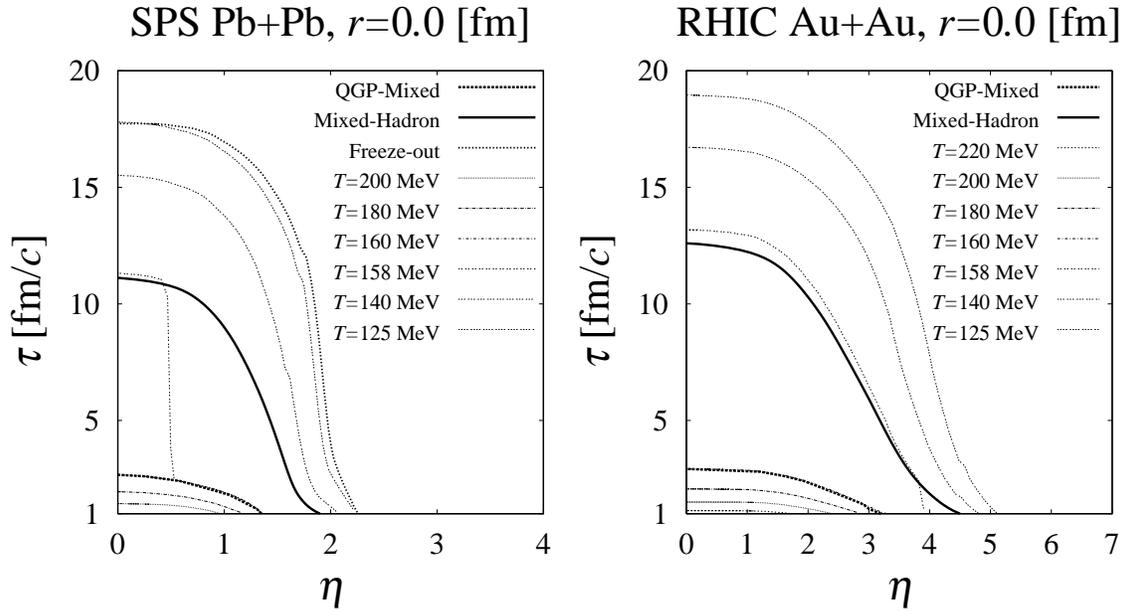}
 \caption{\label{fig:e-t}Temperature contour on $\eta-\tau$ plane at
 $r=0$ fm. Left figure shows the SPS case and right figure shows the
 RHIC case. Thick solid (dashed) line shows the phase boundary between
 mixed (QGP) phase and hadronic (mixed) phase. Freeze-out line is given
 as dotted line for the SPS. As for RHIC, $T_{\text{f}}=125$ MeV contour
 corresponds to the freeze-out line.}
\end{figure}

\begin{figure}[ht]
 \vspace*{1cm}
 \includegraphics[width=0.8\textwidth]{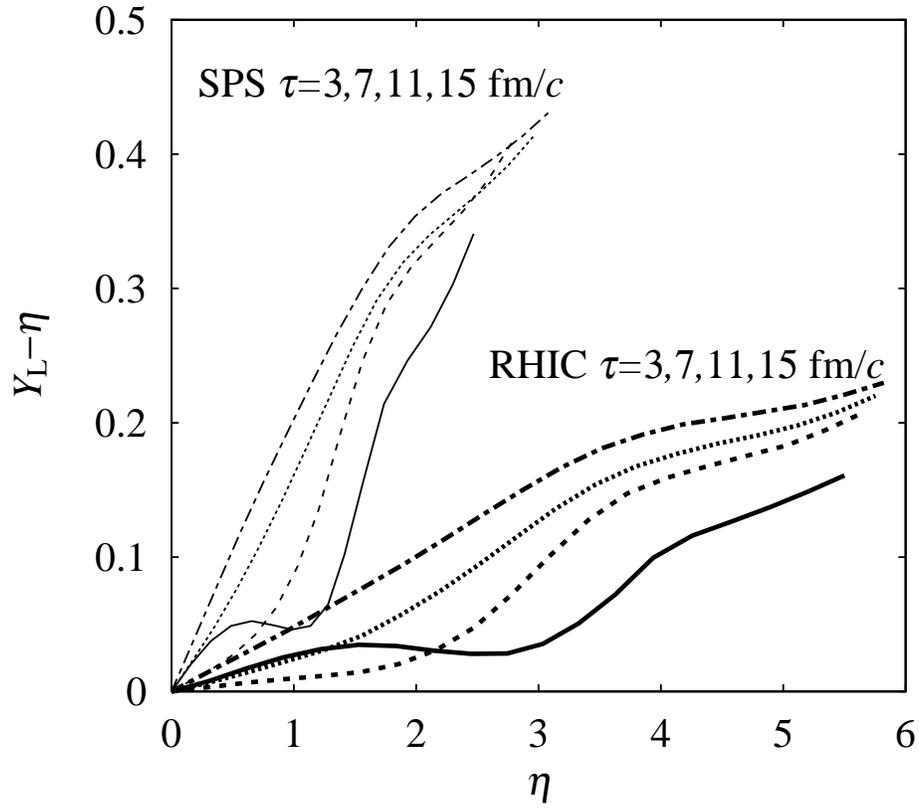}
 \caption{\label{fig:yl}Space-time evolution of longitudinal
 flow at $r=0$. Deviation from Bjorken's scaling solution of
 longitudinal flow rapidity is plotted. Thick lines stand for the RHIC
 case and thin lines stand for the SPS case. In both cases, solid lines,
 dashed lines, dotted lines and dash-dotted lines denote $\tau=$3, 7, 11
 and 15 fm/$c$ cases, respectively.}
\end{figure}

\begin{figure}[ht]
 \includegraphics[height=0.9\textheight]{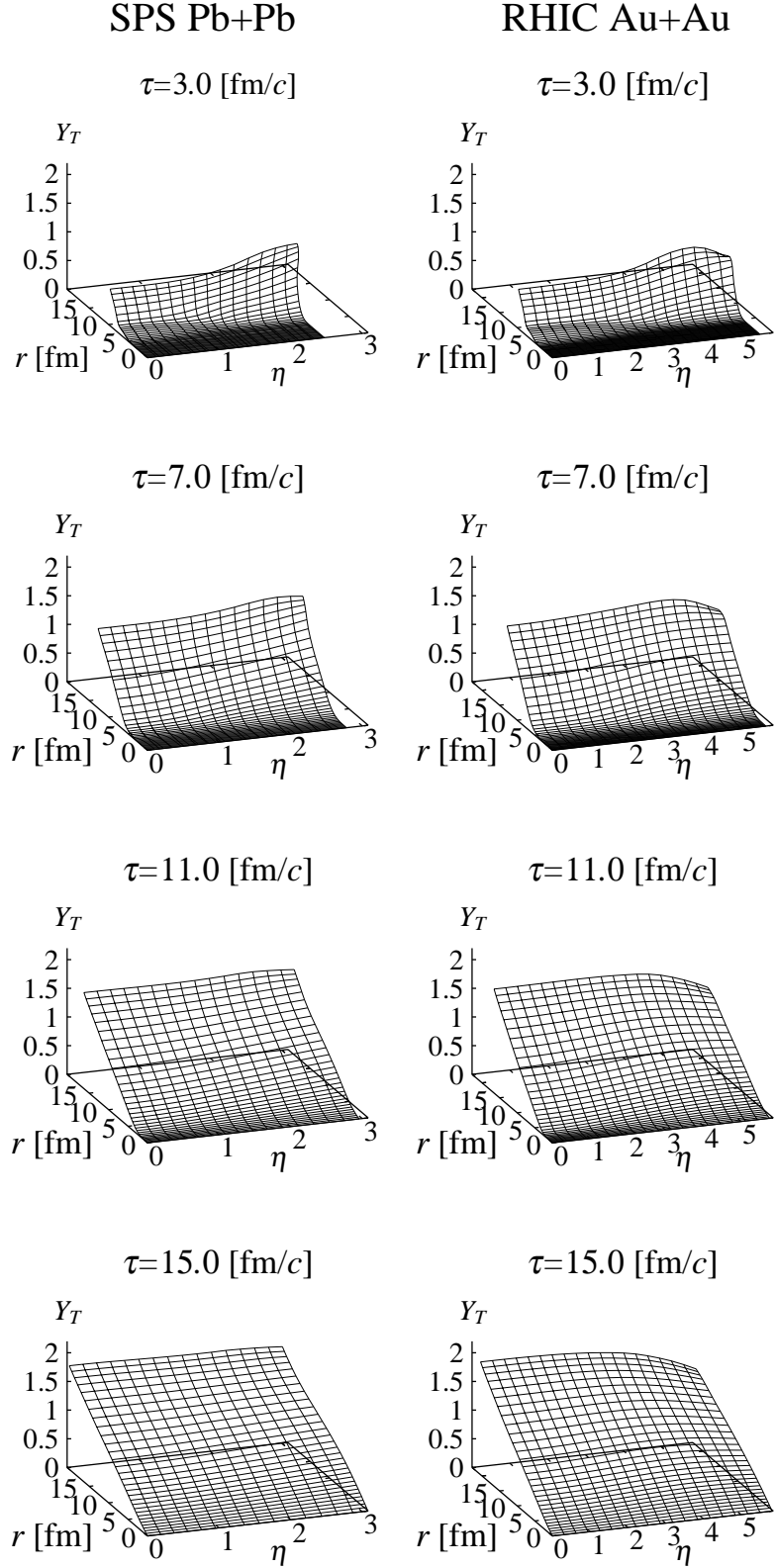}
 \caption{\label{fig:yt}Space-time evolution of transverse flow. Left
 and right column show the SPS and the RHIC cases, respectively.}
\end{figure}

\begin{figure}[ht]
 \includegraphics[width=0.9\textwidth]{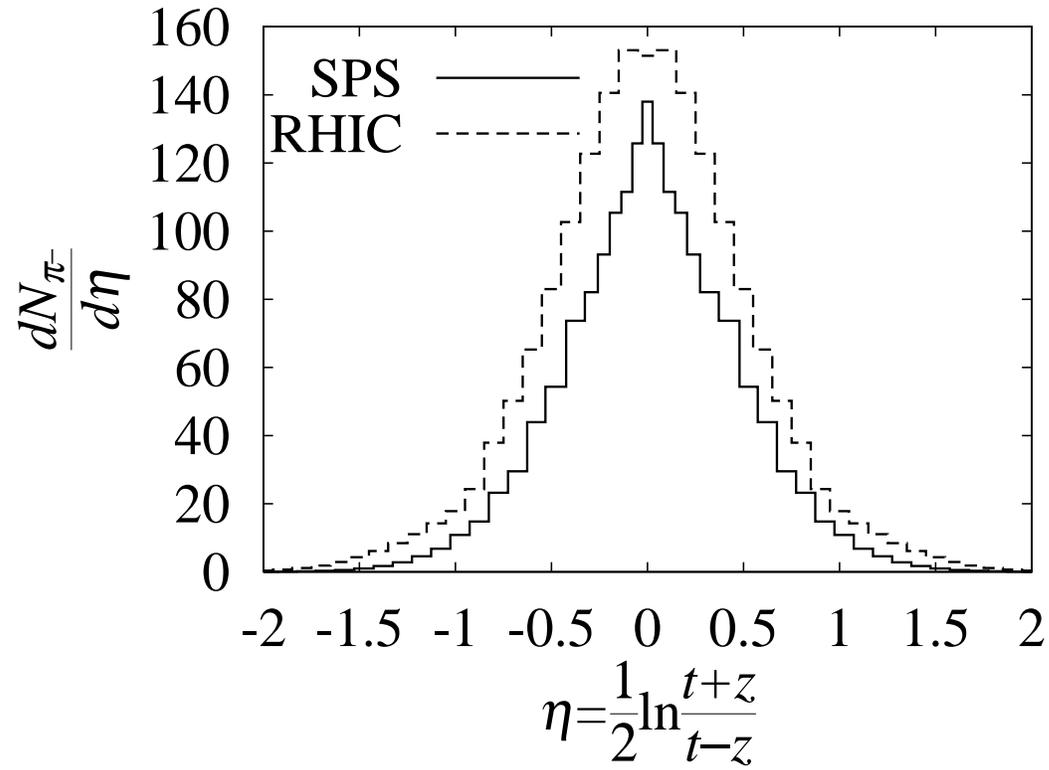}
 \caption{\label{fig:dndetas}Number density of the particles emitted into
 the midrapidity region as a function of space-time rapidity of the
 source point.  Solid line stands for the SPS case and dashed line
 stands for the RHIC case.}
\end{figure}

\begin{figure}[ht]
 \includegraphics[width=0.8\textwidth]{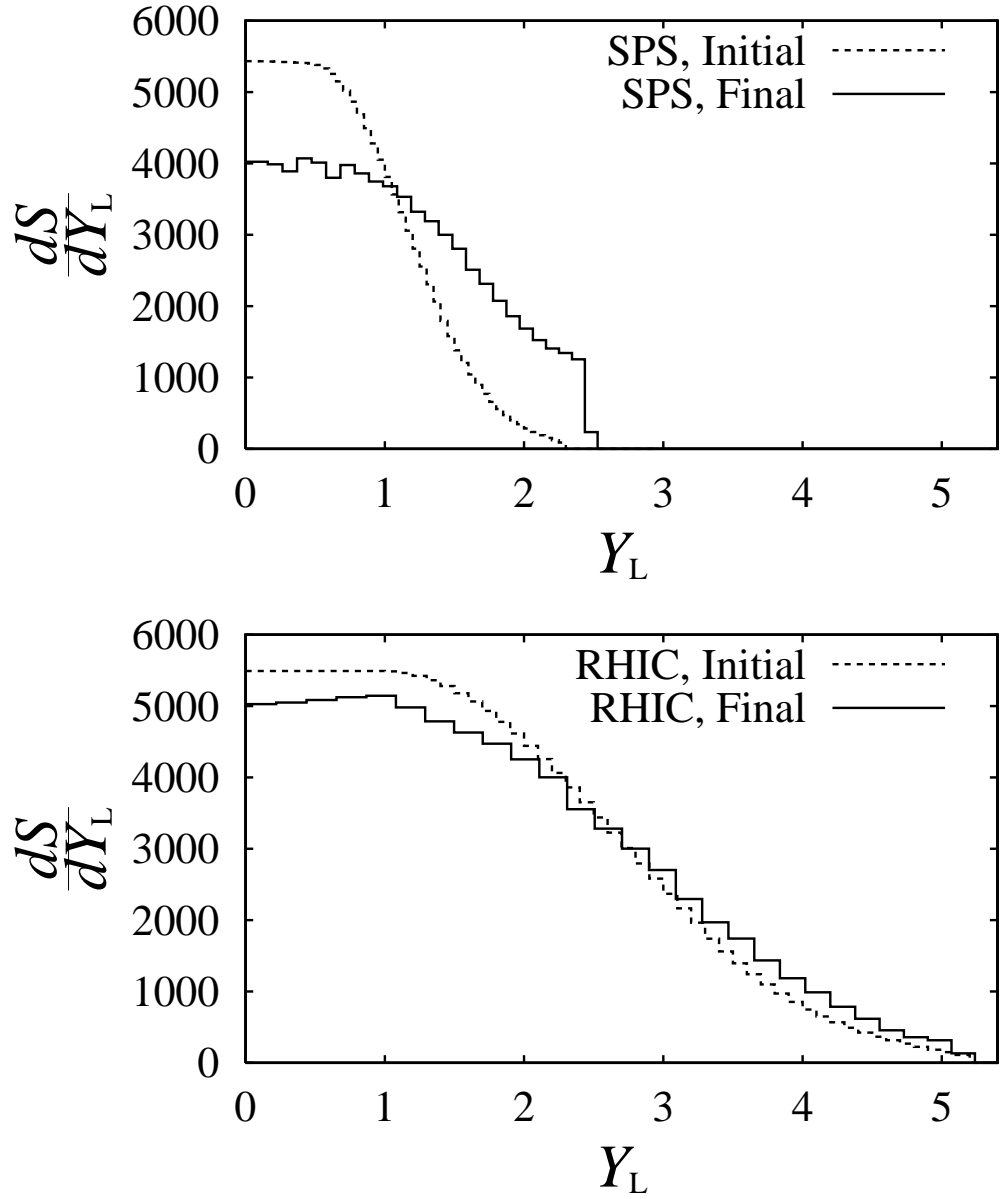}
 \caption{\label{fig:dsdyl}Entropy per unit flow rapidity. In both
 figures (upper for SPS and lower for RHIC), solid lines stand for the
 quantities calculated on
 the freeze-out hypersurface and dashed lines stand for the one on the
 initial stage.}
\end{figure}

In this section, we present the numerical solution of the hydrodynamical
equations and discuss differences in the space-time evolutions of the
fluids between RHIC and SPS. Figures \ref{fig:r-t_sps} and
\ref{fig:r-t_rhic} show space-time evolution of the fluid on the
transverse plane. We also display the space-time evolution on the
longitudinal plane in Fig.~\ref{fig:e-t}. Longitudinal flow and
transverse flow are shown in Figs.~\ref{fig:yl} and \ref{fig:yt},
respectively. From Table \ref{tbl:param}, the maximum energy
density in RHIC is only 5\% higher than the one in SPS. 
Only 5\% higher energy density for the almost 50\% larger $dN/dY$ seems
surprising result. We show the number density of the thermal negative
pions emitted into midrapidty as a function of the space-time rapidity
$\eta$ of the freeze-out point (Fig.~\ref{fig:dndetas}). 
This figure informs us that thermal contribution of the volume element
at $\eta=0$ to the particles into midrapidity is only 9\% larger in RHIC
than in SPS. However, the wider region of the freeze-out hypersurface in
$\eta$ contributes to the midrapidity particle distribution in RHIC more
than in SPS. As a result, factor 1.5 times larger number of particles
obtained in the midrapidity region after summing up particles emitted at
different $\eta$. The difference between RHIC and SPS in
Fig.~\ref{fig:dndetas} originates in the longitudinal extent
$\eta_0+\sigma_\eta$ (see also Fig.~\ref{fig:elong}) and is direct
consequence of longitudinal dynamics. 
We also plot the entropy per unit flow rapidity $dS/dY_{\text{L}}$ in 
Fig.~\ref{fig:dsdyl}. This is a conserved quantity if the boost-invariance
is kept. In both RHIC and SPS, reflecting the deviation from the scaling
solution shown in Fig.~\ref{fig:yl}, entropy is shifted to the larger flow
rapidity. Reduction of entropy at $Y_{\text{L}}=0$, where
$Y_{\text{L}}=\eta=0$ always holds, comes from $dY_{\text{L}}/d\eta$
which is larger than unity \cite{Eskola_EPJC1}. Thus, the shift at RHIC
is smaller than the one at SPS since the deviation from the
boost-invariant solution is small (Fig.~\ref{fig:yl}). At SPS, the
difference of $dS/dY_{\text{L}}$ between the initial stage and the final
stage is larger than the case at RHIC. 

Though our maximum energy density 6.0 GeV/fm$^3$ at initial is also so
smaller than other calculation
\cite{Kolb_PLB500,Zschiesche_nucl7037,Hirano_PRC}, this is
due to the difference of initial time and transverse energy density
profile of which nuclear thickness is considered. As for the average energy
density at midrapidity, we get $\langle \epsilon_{\text{RHIC}} \rangle
=3.9$ GeV/fm$^3$ and $\langle \epsilon_{\text{SPS}} \rangle = 3.77$
GeV/fm$^3$. $\langle \epsilon_{\text{RHIC}} \rangle$ is a little smaller
than an estimation of Ref.~\cite{PHENIX_PRL87}. As a result of such a little
difference in energy density, the space-time evolutions of
two cases do not alter much in Figs.~\ref{fig:r-t_sps} and
\ref{fig:r-t_rhic}. The most different point is a longitudinal extension
of the fluid, $\eta_0+\sigma_\eta$. In RHIC, it is twice as large as in
SPS. This is a consequence of much higher collision energy at
RHIC. Indeed, the total energy of the fluid is 25290 GeV at RHIC, which
is 99\% of total collision energy. Hence, higher collision energy does
not lead to higher energy density but is used to produce the matter with
large volume at $\tau_0 =$1 fm/$c$. 

The output from the fluids is summarized in
Table \ref{tbl:output}. Total net baryon number of the fluid is much
smaller in RHIC than in SPS, as well as the mean chemical potential on
the freeze-out hypersurface. This difference can be seen in the
space-time evolution of temperature on $\eta-\tau$ plane
(Fig.~\ref{fig:e-t}). As shown in Fig.~\ref{fig:phase}, phase boundary
can no longer be specified by temperature only but depends on both
temperature and chemical potential in high net baryon density. For
example, $T=158$ MeV corresponds to the hadronic phase at vanishing baryon
density. However, it can be in QGP phase at $\mu_{\text{B}}=400$
MeV and be in mixed phase at some $\mu_{\text{B}}$. This behavior is
seen $T=158$ MeV contour in Fig.~\ref{fig:e-t}, where the baryonic
chemical potential becomes higher as $\eta$ increases and a fluid
element near $\eta\simeq 0.5$ stays at mixed phase for a long
time. Such behavior does not appear in the RHIC case where chemical
potential is small enough to characterize the mixed phase by the almost
constant temperature. 

In Fig.~\ref{fig:yl} where deviation from Bjorken's scaling
solution $Y_{\text{L}}-\eta$ is plotted, acceleration is larger in SPS
than in RHIC because of the steeper pressure gradient of
$\eta$-direction. Finally, the lifetime of each phase is also shown in
Table ~\ref{tbl:output}. 

 \begin{table}
  \newpage
 \caption{\label{tbl:output}Output quantities from the numerical solutions.}
  \begin{ruledtabular}
   \begin{tabular}[t]{p{4cm}ll}
    & SPS Pb+Pb & RHIC Au+Au \\ \hline 
    Net baryon number & 305 & 131 \\
    Mean freeze-out temperature & 123.2 MeV & 125.0 MeV \\
    Mean chemical potential at freeze-out $\langle \mu_{\textrm{B}} \rangle$ &
    281.6 MeV & 76.1 MeV \\
    Mean transverse flow velocity $\langle v_{\textrm{T}} \rangle $ 
    of the fluid at $|\eta|<0.1$ & 0.467$c$ & 0.509$c$ \\
    Lifetime of the QGP phase $\tau_{\textrm{QGP}}$ & 2.67 fm/$c$ 
    & 2.92 fm/$c$ \\
    Lifetime of the mixed phase $\tau_{\textrm{MIX}}$ & 11.12 fm/$c$ 
    & 12.61 fm/$c$ \\
    Total lifetime of the fluid $\tau_{\textrm{HAD}}$ & 17.74 fm/$c$ 
    & 18.94 fm/$c$ \\
   \end{tabular}
  \end{ruledtabular}
 \end{table}

\section{Two-particle correlation}\label{sec:hbt}
\label{ref}

In this section, we present the result of the
two-pion correlation function and HBT radii based on the numerical
solution of the relativistic hydrodynamical equation. For simplicity, 
we assume that all the pions are emitted from a chaotic source and neglect
the resonance contribution. Then, the two-particle
correlation function is easily calculated through 
\begin{equation}
 C_2(q^\mu, K^\mu)=1+\frac{|I(q^\mu, K^\mu)|^2}{I(0,k_1^\mu)I(0,k_2^\mu)},
  \label{eq:c2}
\end{equation}
where $K^\mu=(k_1^\mu+k_2^\mu)/2$,  $q^\mu = k_1^\mu-k_2^\mu$,
respectively \cite{Shuryak_PLB44, Hama_PRD37}. Here $k_i^\mu$ is
\textit{on-shell} mometum of $i$-th pion. We put
\begin{equation}
 I(q^\mu, K^\mu)= \int K_\tau d\sigma^\tau(x) \sqrt{f(k_1,x)f(k_2,x)}
 \, e^{iq_\nu x^\nu}, \label{eq:interf}
\end{equation} 
so that $I(0,k^\mu)$ reduces to the Cooper-Frye formula with $f(k,x)$ being
the Bose-Einstein distribution function. Considering the experimental
momentum acceptance, we integrate the correlation function with respect
to the average momentum in region $\Omega$ as
\begin{equation}
 C_2(q^\mu)|_\Omega =1+\frac{\int_{\Omega}K_{\text{T}}dK_{\text{T}}dY \,
  |I(q^\mu, K^\mu)|^2}{\int_{\Omega}K_{\text{T}}dK_{\text{T}}dY \,
  I(0,k_1^\mu)I(0,k_2^\mu)}.
  \label{eq:c2int}
\end{equation}
The HBT radii are obtained by fitting the calculated correlation
function \eqref{eq:c2int} to gaussian fitting function
\begin{align}
 C_{\text{2fit}}(q^{\mu})&=1+\exp(-R_{\text{side}}^2 q_{\text{side}}^2
  -R_{\text{out}}^2 q_{\text{out}}^2 -R_{\text{long}}^2
 q_{\text{long}}^2-R_{\text{ol}}^2 q_{\text{out}}q_{\text{long}}).
  \label{eq:c2fit}
\end{align}
For RHIC data, in which rapidity acceptance $|Y|\leq 0.5$, the out-long
cross term $R_{\text{ol}}$ \cite{Chapman_PRL74} can be ignored.
According to the azimuthal symmetry, we can put
$K_{\text{T}}=K_x$, $q_{\text{side}}=q_y$ and $q_{\text{out}}=q_x$.

\begin{figure}[ht]
 \includegraphics[height=0.9\textheight]{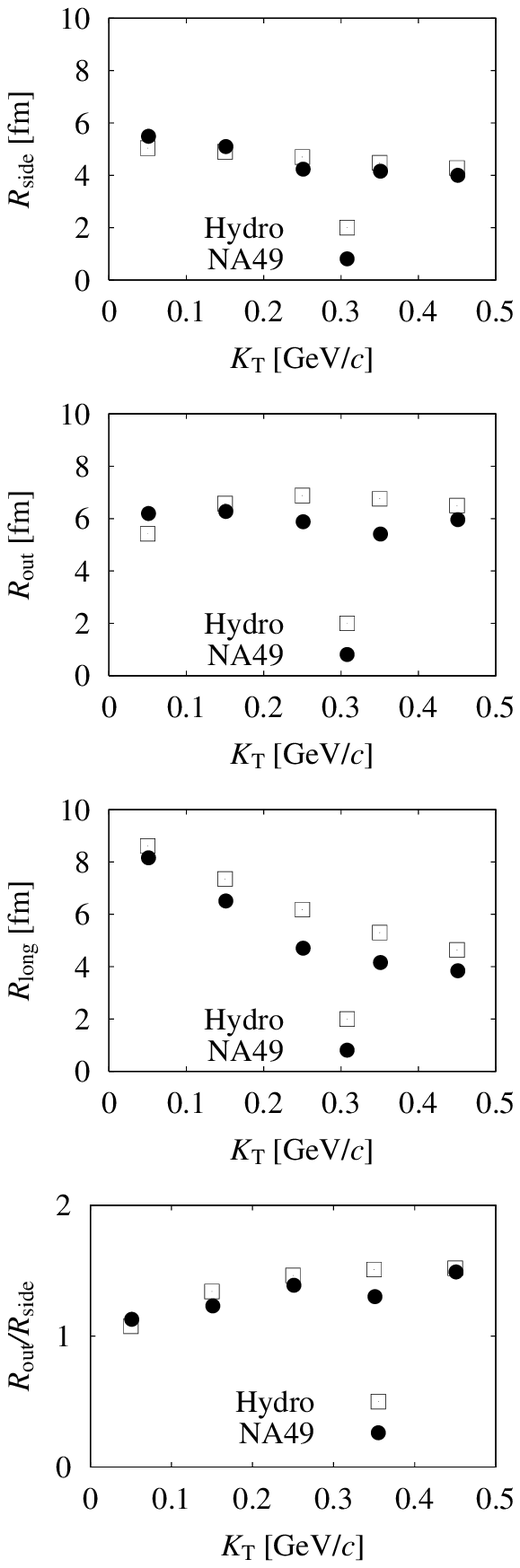}
 \caption{\label{fig:spshbt}HBT radii at SPS. From top to bottom,
 $R_{\text{side}}$, $R_{\text{out}}$, $R_{\text{long}}$ and
 $R_{\text{out}}/R_{\text{side}}$ are plotted. Closed circles denote the
 experimental data from the NA49 collaboration \cite{NA49_QM2001HBT}.
 Open squares stand for our results. Experimental acceptance is 
 $2.9\leq Y \leq 3.4$ in the laboratory system.}
\end{figure}

\begin{figure}[ht]
 \includegraphics[height=0.8\textheight]{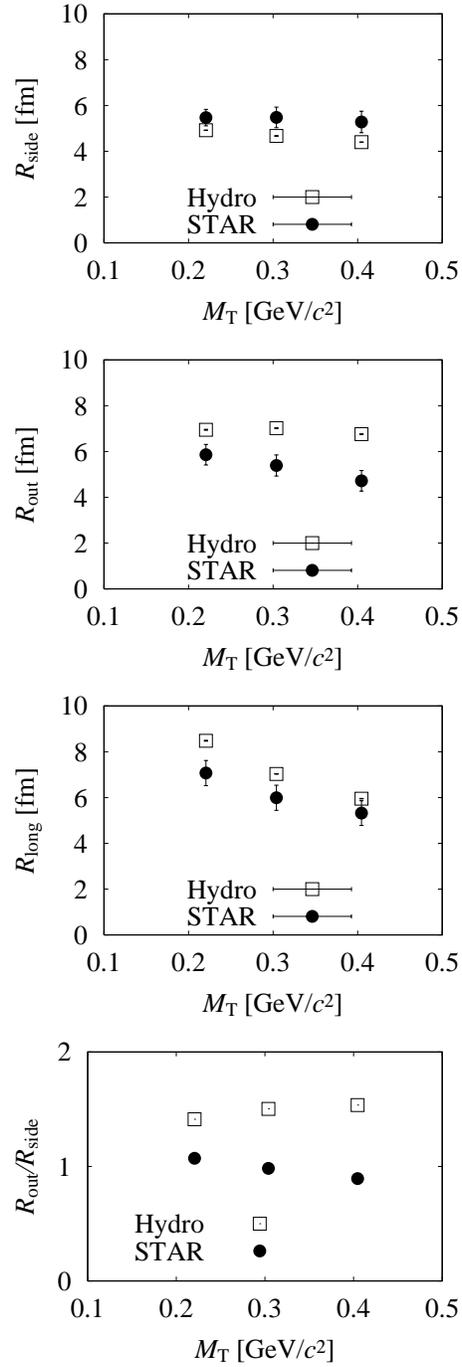}
 \caption{\label{fig:rhichbt}HBT radii at RHIC. From top to bottom,
 $R_{\text{side}}$, $R_{\text{out}}$, $R_{\text{long}}$ and
 $R_{\text{out}}/R_{\text{side}}$ are plotted. Closed circles denote the
 experimental data from the STAR collaboration \cite{STAR_HBT}. Open
 squares stand for our results. Experimental rapidity acceptance is 
 $|Y| \leq 0.5$. Three data points correspond to $0.125 \leq K_{\text{T}} \leq 0.225$ GeV/$c$, $0.225 \leq K_{\text{T}} \leq 0.325$ GeV/$c$ and $0.325 \leq K_{\text{T}} \leq 0.45$ GeV/$c$, respectively.}
\end{figure}

Results of $K_{\text{T}}$ dependence of the HBT radii are
presented in Fig.~\ref{fig:spshbt} and \ref{fig:rhichbt}, where we show the
transverse momentum dependence HBT radii of the SPS Pb+Pb collisions and
$M_{\text{T}}\equiv \sqrt{K_{\text{T}}^2+m_\pi^2}$ dependence of the RHIC
Au+Au collisions, respectively.  In addition to the three radius
parameters, we also present the ratio of $R_{\text{out}}$ to
$R_{\text{side}}$ for better comparison between two collisions
\cite{Rischke_NPA608}. 

Sideward HBT radii (upper figures in Figs.~\ref{fig:spshbt} and
\ref{fig:rhichbt}) are consistent with the experiments in both RHIC and
SPS. Larger radii than other calculations come from initial large
transverse size of the fluid. Outward HBT radii show quantitative
agreement for the SPS data. However, qualitative behavior shows some
deviation from the
experimental data; our result takes the maximum value at
$K_{\text{T}}\simeq 0.3$ GeV/$c$ while the experiment data seem to
monotonically decrease except for the highest $K_{\text{T}}$ bin. 
For the RHIC data, experimental data show steep decrease with
$M_{\text{T}}$. On the other hand, our results are similar to the ones of
SPS because of the similarity in the space-time evolution of both fluids.
As for the longitudinal HBT
radii, our model reproduces the qualitative behavior of the results of
both experiments but shows a little overestimate at low
$M_{\text{T}}$ of RHIC result. Our result
suggests that the longitudinal finite size effect is essential for
understanding the behavior of $R_{\text{long}}$ even at RHIC
because other calculations assuming infinite boost invariant source show
larger $R_{\text{long}}$ \cite{Zschiesche_nucl7037,Heinz_hep0111075}. 
HBT radius in the longitudinal direction has two kinds of origin;
spatial extent of the fluid \cite{Chapman_PRL74} and thermal suppression
caused by rapid expansion. The emission region is roughly
characterized by a product of the Boltzmann factor
$\exp[{-m_{\text{T}}\cosh(Y_{\text{L}}-Y)/T}]$ and
 a geometrical factor (e.g., $\exp(-\eta^2/2(\Delta\eta)^2)$ ).
The deviation from the scaling solution causes stronger thermal
suppression. As a result, our solution provides smaller longitudinal HBT
radius. Our $R_{\text{out}}/R_{\text{side}}$ moderately increases with
$K_{\text{T}}$ in both SPS and RHIC. This tendency is also seen in a
hybrid calculation \cite{Soff_PRL} in spite of the quite
different description of the hadronic phase. Our results show very good
agreement with the SPS result, while the RHIC data clearly shows
different behavior. As concerns the behaviors of $R_{\text{out}}$
and $R_{\text{out}}/R_{\text{side}}$, opaque source
\cite{Heiselberg_EPJC1} is a possible explanation if the current
formalism of the two-particle correlation is correct. Though the opaque
property appears in the hydrodynamical model by virtue of transverse
flow \cite{Morita_PRC}, it is still insufficient to reproduce the small
$R_{\text{out}}$ in the RHIC experiment. More theoretical investigation
will be required for solving the problem
\cite{Humanic_nucl0203004,McLerran_nucl0205028}.

\begin{figure}[ht]
 \includegraphics[height=0.8\textheight]{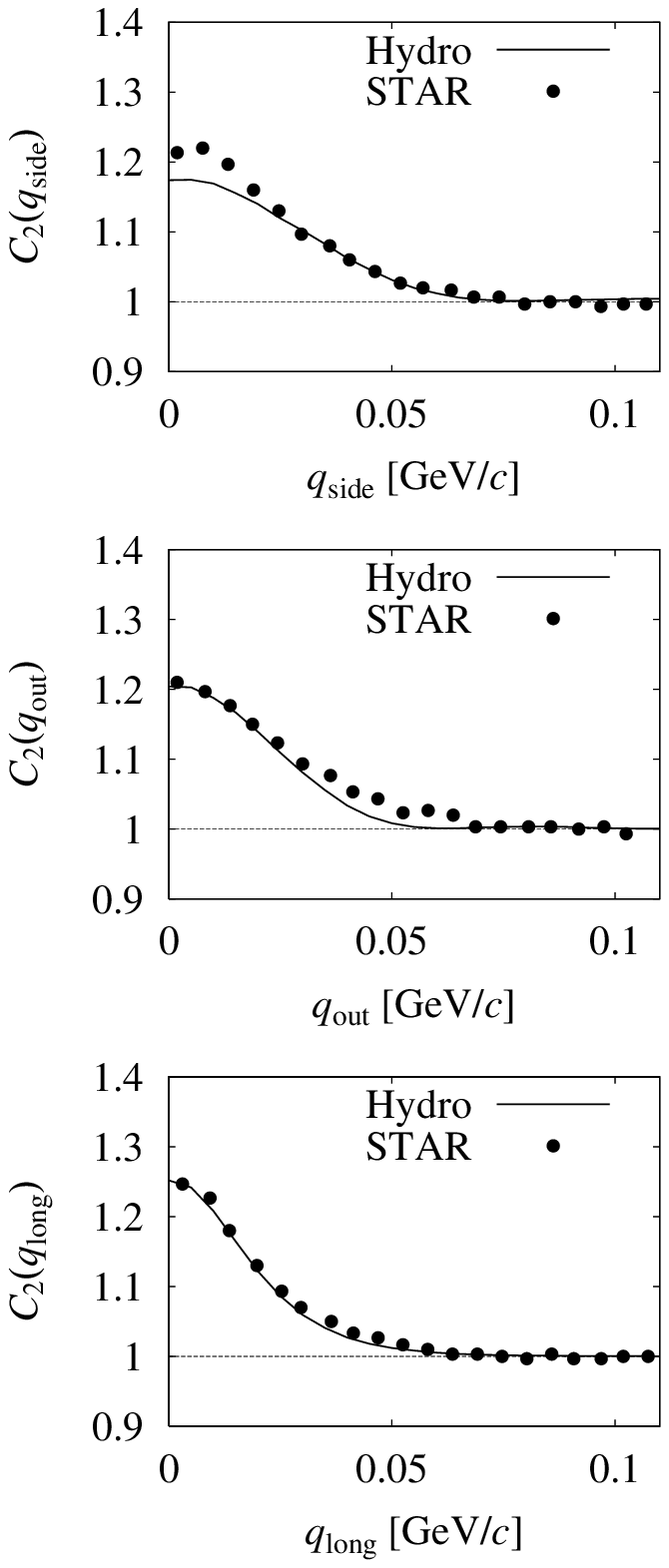}
 \caption{\label{fig:c2}Correlation functions projected onto each
 component of relative momenta. From top to bottom, sideward, outward
 and longitudinal correlation functions are displayed,
 respectively. Each correlation functions are integrated from 0 to 35
 MeV with respect to other two components and corrected by a common
 $\lambda$ factor.}
\end{figure}

Figure \ref{fig:c2} shows the correlation functions projected onto each
components of relative momenta. Transverse momentum of emitted pair is
$0.125 \leq K_{\text{T}} \leq 0.225$ GeV/$c$, which corresponds to the lowest
momentum bin in Fig.~\ref{fig:rhichbt}. In each figure, calculated
correlation functions are corrected by a common factor as
$\lambda=0.6$. The other kind of the reduction of the correlation
function at $q_i =0$ ($i=$side, out, long) is caused by an integration
with respect to other components of the relative momenta. The resultant
factor is in proportional to $1/R_i$. Our correlation function seems to
be consistent with the experimental data for the outward and the
longitudinal direction in spite of the overestimation of the HBT
radii. Because the HBT radii correspond to the inverse width of the
correlation function, difference in width between $R=7$ fm and $R=6$ fm
is only about 5 MeV in relative momentum. The small value of sideward
correlation function at $q_{\text{side}}=0$ indicates that outward and
longitudinal HBT radii are larger than the experimental results.

\section{Concluding Remarks}\label{sec:conclusion}

In this paper, we investigate single-particle distributions and
two-particle correlation functions in SPS Pb+Pb 17 $A$GeV collisions and
RHIC Au+Au 130 $A$GeV collisions based on a hydrodynamical model in
which both longitudinal and transverse expansion are taken into
account. As long as the single-particle spectra, the hydrodynamical
model well describes both SPS and RHIC data. The initial parameter set
in the model for both collisions indicates that initial energy density
in RHIC is only slightly higher than the one in SPS and much larger
extent of hot matter is produced in RHIC, if we compare them at the same
initial time and by similar initialization (Fig.~\ref{fig:initial} and
Eqs.~\eqref{eq:eini},\eqref{eq:nbini}). We have also discussed the
space-time evolution of the
fluids. Since the initial conditions are not different so much, 
temperature and transverse flow evolution do not show significant
difference between SPS and RHIC.  Only the equi-temperature contour of $T=158$
MeV shows the qualitative difference between SPS and RHIC due to the
difference in net baryon number.  
Steeper pressure gradient in the longitudinal direction at SPS makes
the deviation from the scaling solution larger than at RHIC.
Two-pion correlation functions and the
HBT radii are also investigated. Our model shows good agreement with the
SPS data. For RHIC data, however, the outward and longitudinal HBT radii
of our result are larger than the experimental data even though the
dynamical longitudinal expansion and finite size effect are incorporated. 

\begin{acknowledgments}
 The authors would like to thank Prof. I.~Ohba and Prof. H.~Nakazato for their
 fruitful discussions and comments. They also would like to acknowledge
 J.~Alam, R.S.~Bhalerao, P.~Kolb, L.~McLerran, H.~Nakamura, and
 F.~Retri\`{e}re for valuable discussions and comments. This work was in part
 supported by the Ministry of Education, Science and Culture, Japan
 (Grant No.~11440080 and 13135221), Waseda University Grant for Special
 Research Projects No.2001A-888, Waseda University Media Network
 Center and ERI of Tokuyama University.
\end{acknowledgments}

\end{document}